\documentclass{ws-procs975x65}

\def\beq{\begin{equation}}
\def\eeq{\end{equation}}

\begin{document}

\title{Causal Set theory, non-locality\\ and phenomenology}
\author{Alessio Belenchia}

\address{SISSA,\\
SISSA - International School for Advanced Studies, Via Bonomea 265, 34136 Trieste, Italy\\
INFN, Sezione di Trieste, Trieste, Italy\\
$^*$E-mail: abelen@sissa.it
}



\begin{abstract}
This proceeding is based on a talk prepared for the XIV Marcel Grossmann meeting. We review some results on causal set inspired non-local theories as well as work in progress concerning their phenomenology.\\

\end{abstract}


\bodymatter


\section{Introduction}

Quantizing gravity and/or the quantum nature of spacetime have been one of the central themes in theoretical physics in the last century. The two pillars of modern physics, i.e. general relativity and quantum theory are reluctant to be reconciled despite many efforts. In spite of many theoretical advancements the field of quantum gravity (QG) still lacks contact with observations and experiments. However, in recent years the field of quantum gravity phenomenology has started to attract the interest of the community. Since to date we do not have a fully working quantum gravity model, but a plethora of models each with nice features and drawbacks, we should bet on common features shared by different models. 
One of the recurring features in models of QG is the idea that spacetime may be fundamentally discrete.
This idea is common to various approaches even if it presents itself in different ways .~\cite{Garay:1994en} However, when dealing with discrete spacetime, Lorentz Invariance violating (LIV) effects are lurking. Indeed studies of LIV are a good example of QG phenomenology that has flourished in the last 20 years or so. We nowadays have stringent constraints on LIV coming from experiments ranging from astrophysical observations to earth based high energy experiments. Together with theoretical arguments, these constraints tightly restrict the number of viable LIV theories and suggest that Lorentz Invariance (LI) may actually be an exact symmetry of Nature telling us something deep about spacetime. From these latter considerations follows the idea that inspired our investigation: to consider discrete spacetime models that preserve Lorentz Invariance and study their phenomenology.        

In this work we consider a model of discrete spacetime, causal set theory (CS), which is LI and its associated phenomenology. We will also propose a way to test non-local features, results of the discreteness itself, with state of the art (or near future) quantum experiments.

The paper is organized as follows. In section~\ref{sec2} we briefly introduce the basic principles of causal set theory.
In section~\ref{sec3} we explore Huygens' principle violations in flat 4 dimensional spacetime and  possible dimensional reduction in causal set theory. In section~\ref{sec4} we propose a way to test non-local effects using opto-mechanical quantum oscillators. We conclude in section~\ref{sec5} with a discussion and future directions.    

\section{Causal set, non-locality and phenomenology}\label{sec2}

Causal set theory is an approach to quantum gravity that stems from the assumptions that the causal structure is fundamental and spacetime is discrete. In particular a causal set is defined as a partially ordered, locally finite set.~\cite{Surya:2011yh}  
  
Already in general relativity (GR) spacetime is a partially ordered set, where the partial order represents the causal structure, but not a locally finite one. Moreover, it is striking that the knowledge of solely the causal structure of spacetime can be proved to be enough, under some technical causality conditions, to completely determine the conformal geometry (see Ref.~\refcite{Surya:2011yh} and reference therein). These results in differential geometry shine a light on the primitive importance of causal structure, which is assumed to be fundamental in CS theory. Finally, the locally finite requirement translates into rigorous terms the physical expectation that spacetime is discrete at scales comparable with the Planck scale as suggested by different theoretical hints.~\cite{Sorkin:1997gi}

In what follows we will be interested in the kinematic structure of causal sets which offers  interesting phenomenological possibilities for what concerns the study of the effects of the spacetime discreteness. Indeed, different phenomenological models and predictions have been produced in the CS literature, ranging from the prediction of the cosmological constant to swerves model.~\cite{Sorkin:1990bh,Dowker:2003hb} In particular, we will consider causal sets that well approximate flat spacetime in different dimensions. These causal sets are the ones that can arise with relatively high probability from the Poisson process called \textit{sprinkling}. This process consists in selecting points uniformly at random in Minkowski with density $\rho= l^{-d}$ (being $l$ the fundamental discreteness scale and $d$ the spacetime dimension) so that the expected number of points sprinkled in a region of spacetime volume $V$ is $\rho V$. The sprinkling process produce a random lattice that preserve Lorentz invariance in a well defined sense, in this way we have a kinematical randomness that preserves Lorentz invariance while working with a discrete structure.~\cite{Bombelli:2006nm} However, there is a price to pay: A fundamental non-locality of causal sets. Indeed, consider the nearest neighbours to a given point in a causal set well-approximated by Minkowski spacetime. These will lie roughly on the hyperboloid lying one Planck unit of proper time away from that point and therefore will be infinite in number. This non-locality manifests itself also in the definition of the non-local d'Alembertian for a scalar field on the causal set. ~\cite{Sorkin:2007qi} The non-local d'Alembertian is a discrete operator that reduces in the continuum limit to the standard (local) wave operator. The precise form of this
correspondence is given by performing an average of the causal set d'Alembertian over all
sprinklings of Minkowski, giving rise to a non-local, retarded, Lorentz invariant
linear operator in the continuum, $\Box_{nl}$, whose non-locality is parametrised by a scale $l_{k}$\footnote{This scale should be much larger than the discreteness one in order to tame the fluctuations of the discrete operator, see Ref.~\refcite{Sorkin:2007qi}.}. Locality is restored in the limit $l_k \rightarrow 0$ in which $\Box_{nl}\rightarrow\Box$. We will consider the non-locality scale $l_{k}$ as a free parameter of the theory on which we would like to cast phenomenological bounds.

The general expression for the non-local d'Alembertians in flat spacetime of dimension $d$ was introduced in Ref.~\refcite{Aslanbeigi:2014zva} and is given by
\begin{equation}
\Box_{\rho}^{(d)}\phi(x) = \rho^{\frac{2}{d}}\left(\alpha^{(d)}\phi(x) +\rho\, \beta^{(d)}\sum_{n=0}^{N_d}C_n^{(d)}\int_{J^-(x)}d^dy \,\frac{(\rho V(x,y))^n}{n!}e^{-\rho V(x,y)}\phi(y)\right),
\label{boxd}
\end{equation}
where $N_{d}$ is a dimension dependent positive integer, $\rho=1/l_k^d$ and the coefficients $\alpha^{(d)}$, $\beta^{(d)}$ and $C_n^{(d)}$ can be 
found in Equations (12)-(15) of Ref.~\refcite{Dowker:2013vba}. In Ref.~\refcite{Belenchia:2014fda} the canonical quantization of the non-local, free scalar field theory was performed. A different quantization scheme was investigated, considering non-local field theories that share some common features with the ones deriving from causal sets, with similar results.~\cite{Saravani:2015rva}

\section{Huygens' principle violations and dimensional reduction}\label{sec3}          
 In the following we consider two features that emerge from the study of Eq.~\eqref{boxd}, i.e. Huygens principle violations in 4 dimensional flat spacetime and dimensional reduction. Where the first can be conceived as pointing towards interesting phenomenological scenario the second has intrinsic theoretical importance since dimensional reduction is a common features of different QG approaches.

\subsection{Huygens' principle}   
\begin{figure}[h]
\begin{center}
\includegraphics[scale=0.2
]{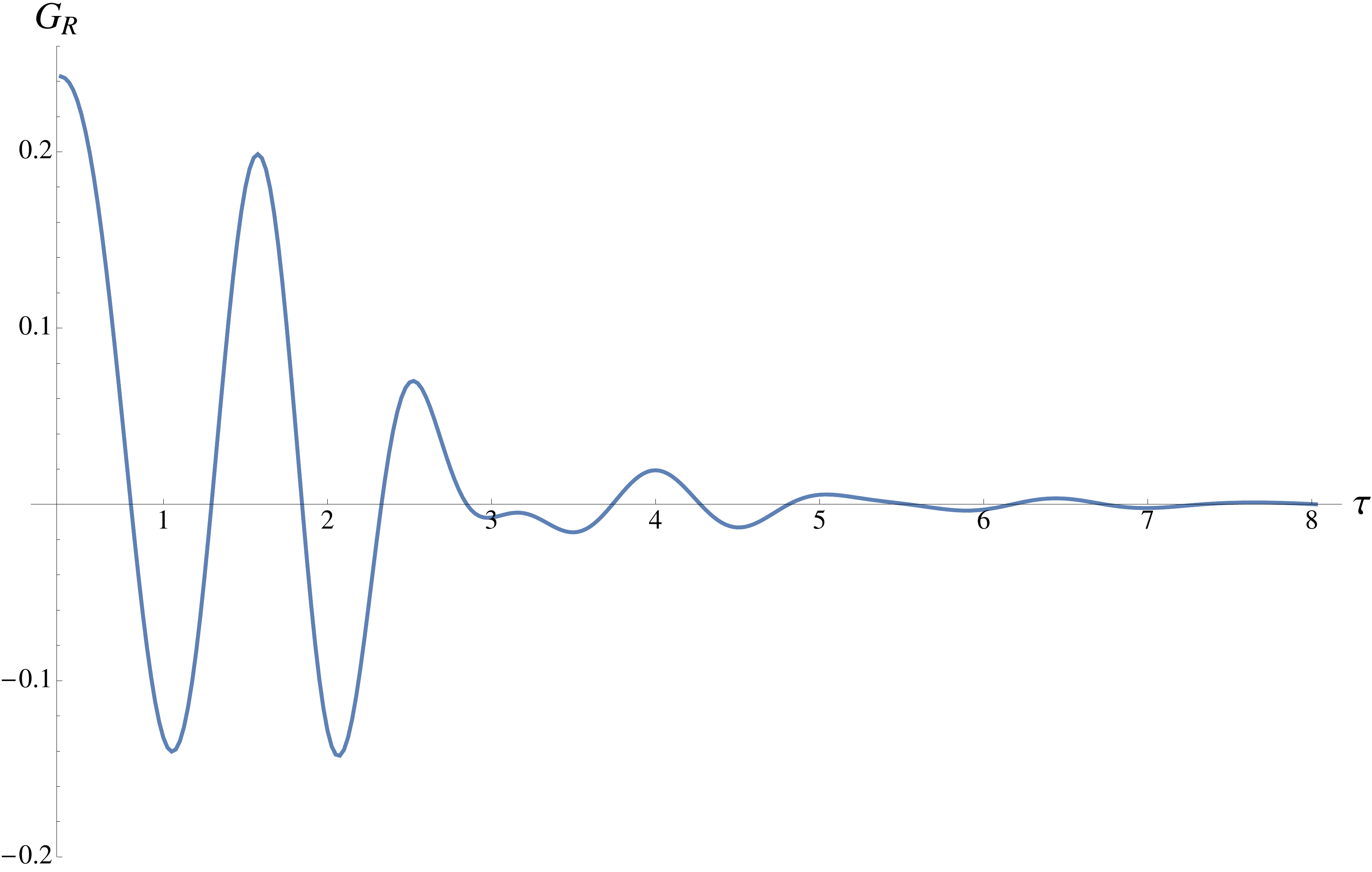}
\end{center}
\caption{Retarded Green function of the non-local d'Alembertian in 4 dimensions as a function of the proper time $\tau$.}
\label{gr}
\end{figure}

Huygens' principle (HP) states that in spacetime dimension $d =2n+2$, $n > 0$, the Green functions of the wave equation have support on the light-cone,
while in dimensions d = 2n + 1 they also have support inside the light-cone. \footnote{The case d = 2
is degenerate since the Green functions are constant inside the light cone.} This translates into the fact that the field perturbations can propagate only on the light cone when HP is satisfied. The fact that HP violations are present in 4-dimensional flat spacetime for the non-local causal set d'Alembertian was observed in Ref.~\refcite{Belenchia:2014fda}\footnote{See also Ref.~\refcite{Johnston:2010su}.} by looking at the support of the retarded Green function, see Fig.~\ref{gr}.
It has been recently shown that, whenever HP is violated\footnote{That is the case also in presence of curvature in general, so that effects related to HP violations could be important in cosmological scenarios.} an information channel opens that permits information transmission without energy exchange.~\cite{Jonsson:2014lja} It is then interesting to observe that, the causal set non-local d'Alembertian shows HP violations in flat 4-dimensional spacetime. Indeed, if it is possible to extend such results to the electromagnetic field this will open to the possibility of testing HP violations in the lab using high precision experiment to cast bounds on non-local effects deriving from CS discreteness.

\subsection{Spectral dimension}
\begin{figure}[h]
\centering
\includegraphics[scale=0.2]{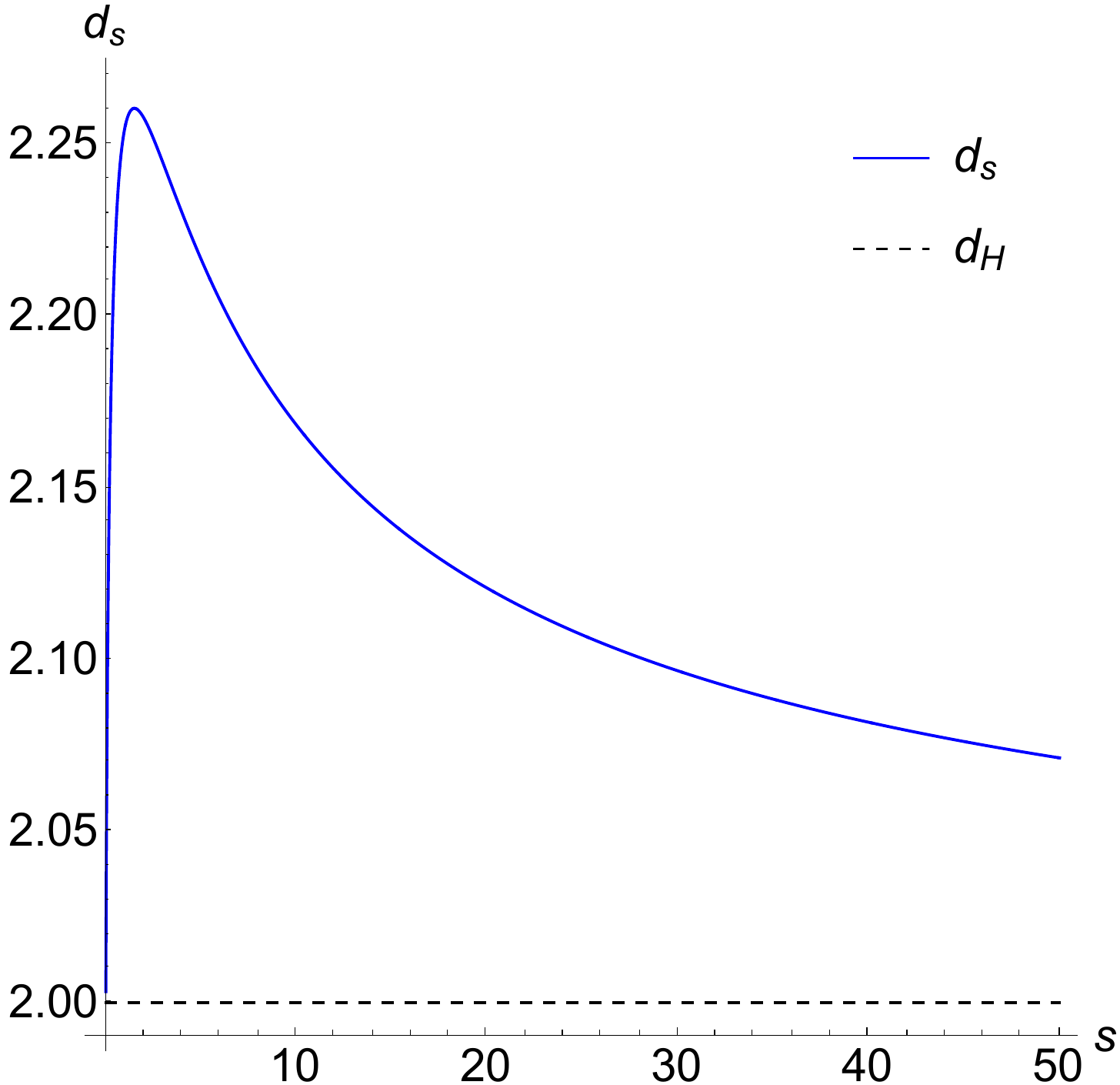}
\includegraphics[scale=0.2]{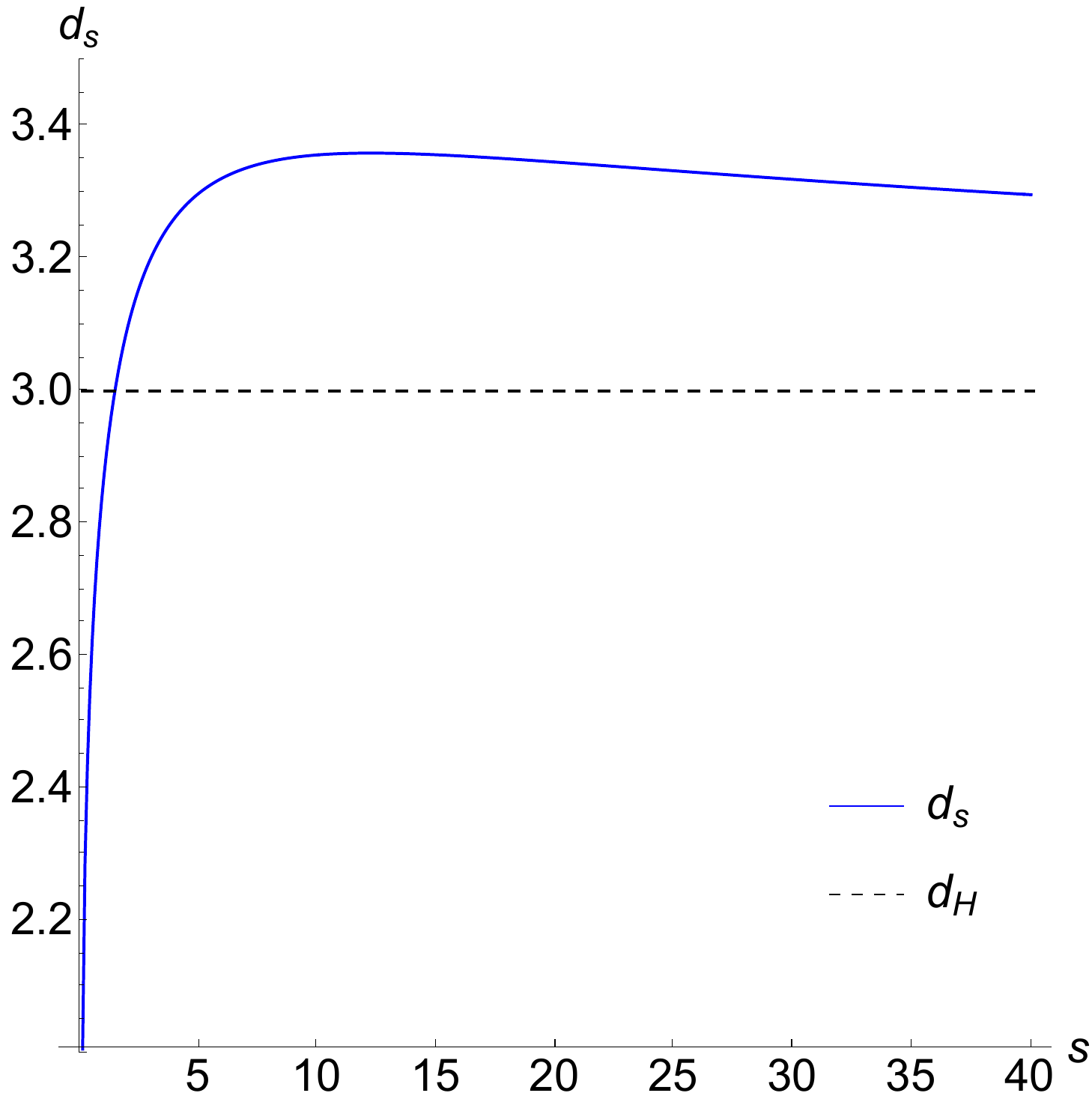}
\includegraphics[scale=0.2]{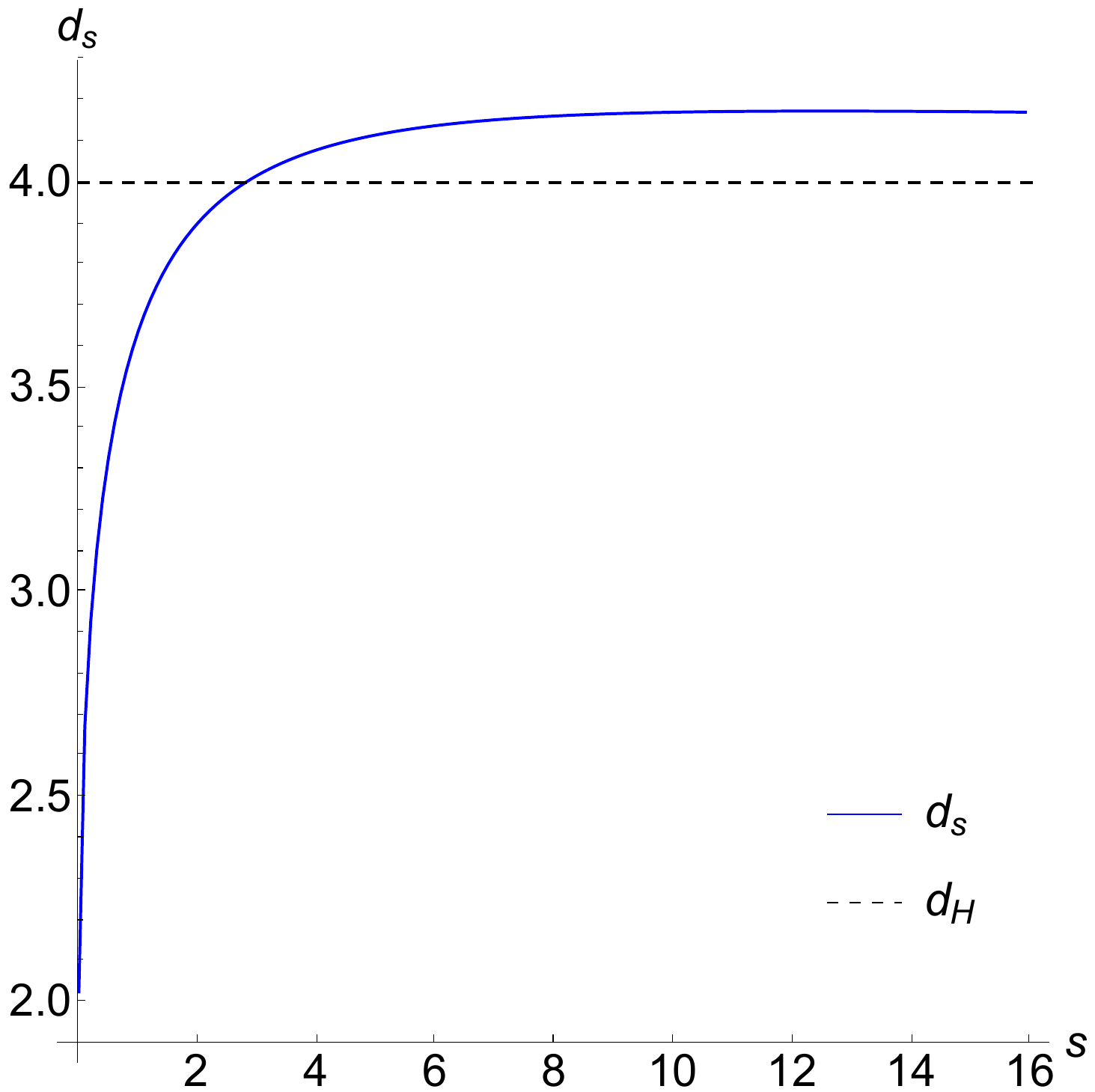}
\caption{From left to right we have the spectral dimension $d_s$ as a function of diffusion time $s$ for $\rho=1$ in dimension $d=2,3$ and 4 respectively. The dashed line represents the value of the Hausdorff dimension ($d$). The spectral dimension interpolate between $d_s=2$ at short scales and $d$ at large scales. 
}\label{sd}
\end{figure}

Spectral dimension in CS theory was first computed in Ref.~\refcite{Eichhorn:2013ova} using a random walker on the graph. The result was that the spectral dimension, computed in this way, diverges in the limit of short diffusion times. This behaviour was explained using the non-local nature of causal sets discussed previously. However, we computed the spectral dimension associated to the non-local scalar field d'Alembertian finding that, for every dimension, in the UV there is a dimensional reduction to 2 of the spectral dimension, see Fig.~\ref{sd}.~\cite{Belenchia:2015aia} 
To date the precise link between our result and the one of Ref.~\refcite{Eichhorn:2013ova} has not been clarified. It is possible that the two quantities carry different physical meaning given that they are obtained with different methods and in different regimes of the theory (albeit having the same name), or that the spectral dimension computed in Ref.~\refcite{Belenchia:2015aia} ceases to be valid at scales when the continuum approximation is no longer reliable. Note moreover that additional hints towards dimensional reduction in causal set theory were recently presented.~\cite{Carlip:2015mra}    

\section{Quantum mechanical test of non-locality}\label{sec4}
In this last section we consider a recently proposed way to test models that present non-locality\footnote{These are not limited to CS inspired models and, moreover, the present analysis is not directly applicable to the CS case.} with opto-mechanical experiments.~\cite{Belenchia:2015ake} The rationale behind this  proposal rests in the fact that quantum mechanical experiments are reaching regimes in which it is possible to test some general ideas stemming from QG, see e.g. Ref~\refcite{Bawaj:2014cda}.

\begin{figure}[h]
\begin{center}
\includegraphics[scale=0.4
]{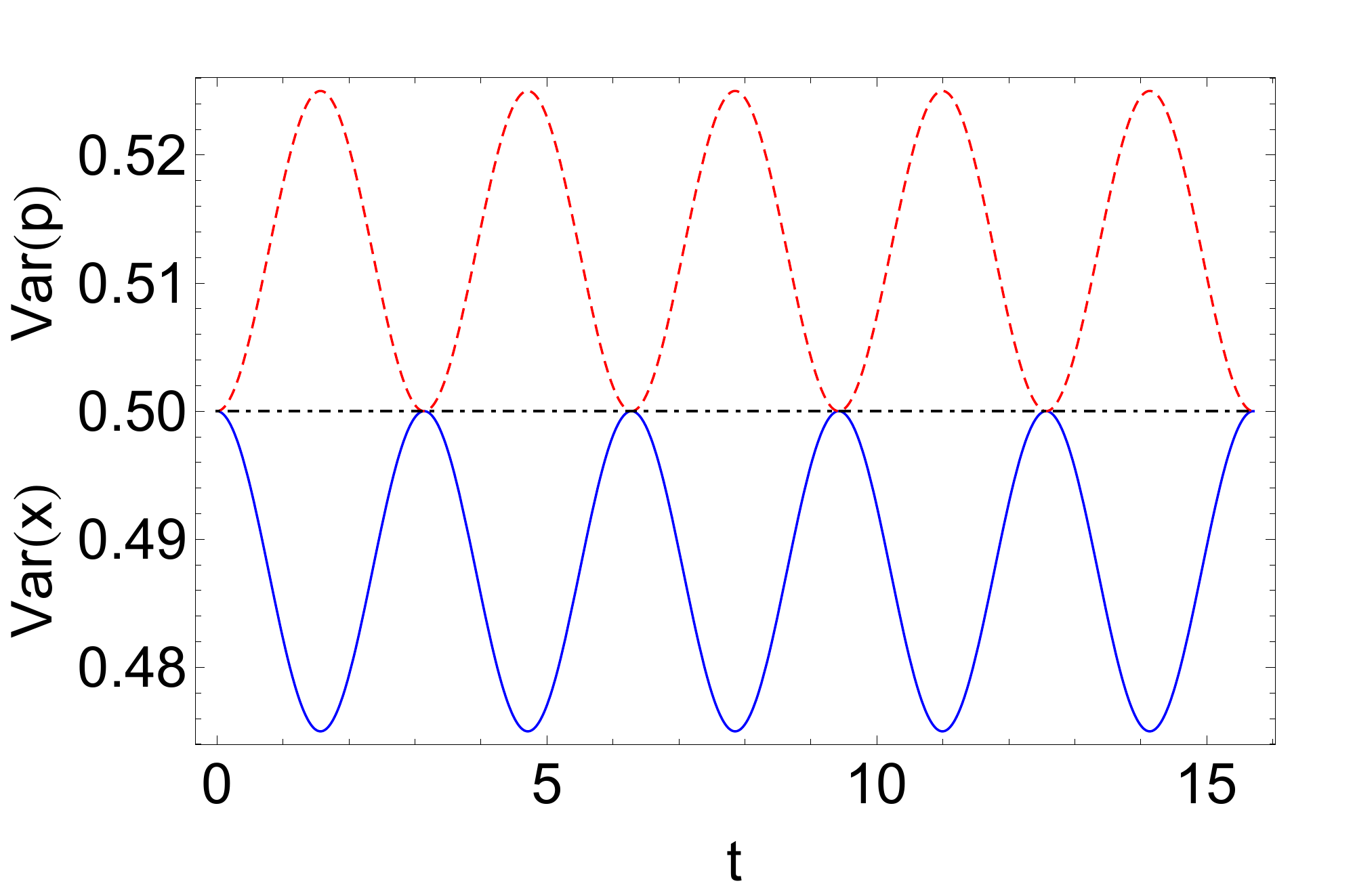}
\end{center}
\caption{Time dependence of the variances of a coherent state with $\epsilon$ exaggerated to be $\approx \mathcal{O}(10^{-2})$. The continuous (blue) and dashed (red) lines represent the position and momentum variances respectively. The black dot-dashed line is the local value, 1/2.}
\label{nls}
\end{figure}

We studied the effect of non-locality in the evolution of coherent states of a quantum harmonic oscillator, a system that is central in opto-mechanical experiments. In particular we derived, starting from a non-local Klein-Gordon equation of the type $f(\Box-m^{2})$ with $f$ an analytic function, the corresponding non-local Schr\"odinger equation. We then solved this equation perturbatively around a coherent state solution of the local Schr\"odinger equation, the perturbation parameter being $\epsilon\equiv m\omega l_k^2/\hbar$ (where $m$ is the mass of the system and $\omega$ its natural frequency). What we found is that the first order correction to the coherent state evolution produce a \textit{spontaneous, time periodic squeezing} of the state and, indeed, the state remain of minimum uncertainty throughout its evolution, see Fig.~\ref{nls}.

We then used current data coming from opto-mechanical experiments to forecast the expected bounds on the non-locality scale that could be achieved with ongoing experiments. The forecast shows how very stringent bounds will possibly be achieved, of the order of $l_{k}\leq 10^{-29}$m. Even tough six order of magnitude from the Planck scale these kind of bounds can explore a very interesting and uncharted territory, far beyond current bounds coming from LHC data $l_{k}\leq 10^{-18}$m~\cite{Biswas:2014yia}.

\section{Conclusion and future directions}\label{sec5}   
Quantum gravity phenomenology is attracting more and more interest in recent years. In this work we have reviewed the phenomenological possibilities related to non-local d'Alembertians in causal set theory as well as a new proposal for testing non-local models with opto-mechanical experiments. Huygens' principle violations seems to have the potential for be used to constrain non-local models stemming from CS theory, work in this direction is in progress. On the other hand, tests of (quantum) gravity inspired effects with ``macroscopic'' quantum objects are being studied in different contexts and promise to shed new light on our ides about space, time and gravity.

\section*{Acknowledgments}
The author would like to thank Dionigi Benincasa for comments and suggestions on early drafts and Stefano Liberati for illuminating discussions and support. We wish to acknowledge the John Templeton Foundation for the supporting grant \#51876.

\end{document}